\documentclass[%
 reprint,
 amsmath,amssymb,
 aps,
nofootinbib
]{revtex4-2}

\usepackage{revquantum}
\usepackage[super]{nth}% To be able to write 1st, 2nd, etc.
\usepackage{graphicx}% Include figure files
\usepackage{dcolumn}% Align table columns on decimal point
\usepackage{bm}% bold math
\usepackage{upgreek}% Non-italic Greek letters
\usepackage{xcolor}% Used to set colors to, e.g., text (comments to myself or others). 
\usepackage{enumitem}
\usepackage{placeins}% Used for FloatBarrier
\usepackage{braket}
\usepackage{siunitx}
\newcommand{\red}[1]{\textcolor{black}{#1}}

\usepackage{dsfont}
\usepackage{comment}

\begin{document}

\title{Complete analysis of a realistic fiber-based quantum repeater scheme}

\author{Adam Kinos}
\email{adam.kinos@fysik.lth.se}
\author{Andreas Walther}
\author{Stefan Kr\"{o}ll}
\author{Lars Rippe}
\address{Department of Physics, Lund University, P.O. Box 118, SE-22100 Lund, Sweden}

\date{\today}

\begin{abstract}
\red{We present a quantum repeater protocol for distributing entanglement over long distances, where a dedicated communication stage enables trial rates not limited by the travel time between repeater nodes. To accomplish this, each node contains several qubits that can couple to one single-photon emitter. Photons from the emitters generate heralded entanglement between qubits in neighboring nodes. The protocol leaves the emitters disentangled from the rest of the system immediately after emitting the photons, thus allowing them to be reused to entangle other qubits without waiting for the repeater link round-trip time. This time multiplexing increases the protocol trial rate by up to an order of magnitude. The protocol is then combined with conventional deterministic entanglement swapping and heralded entanglement purification to extend the entanglement distance and reduce the entanglement error, respectively.} We perform a complete protocol analysis by considering all relevant error sources, such as initialization, two-qubit gate, and qubit measurement errors, as well as the exponential decoherence of the qubits with time. The latter is particularly important since we analyze the protocol performance for a broad range of experimental parameters and obtain secret key rates ranging from $1 \rightarrow 1000$ Hz at a distance of $1000$ km. Our results suggest that it is important to reach a qubit memory coherence time of around one second, and two-qubit gate and measurement errors in the order of $10^{-3}$ to obtain reasonable secret key rates over distances longer than achievable with direct transmission. \red{While this work focuses on optimizing secret key rates, the protocol can also be used for EPR pair generation and is thus also relevant for, e.g., distributed quantum computing.}
\end{abstract}

\maketitle

% ---------------------------------------------------------------
\section{\label{sec:intro}Introduction}
Quantum repeaters are essential for establishing long-distance quantum networks since direct transmission suffers from exponential losses with distance \cite{Takeoka2014, Muralidharan2016}. These losses can be overcome by breaking the total distance into shorter links and using heralded entanglement generation (HEG) and entanglement swapping to obtain EPR pairs (Bell states) between the end nodes. Quantum repeater protocols using atomic ensembles and linear optics have been suggested \cite{Duan2001, Sangouard2011, Bussieres2013}, and recent overviews of the present status can be found in Refs. \cite{Wu2020, Lei2023}. However, such protocols often rely on probabilistic entanglement swapping, which has a detrimental effect on longer quantum networks. To overcome this problem, single-photon-emitter protocols using deterministic entanglement swapping based on two-qubit gates (TQG) and qubit measurements have been proposed \cite{Childress2006, Sangouard2009, Asadi2018}. Furthermore, to limit the errors of the EPR pairs, one can use heralded entanglement purification protocols that probabilistically convert two EPR pairs into one with a lower error \cite{Deutsch1996, Dur1999, Krastanov2019}. Alternatively, error correction codes can be used \cite{Jiang2009, Munro2010, Zwerger2018}, but those usually require more qubits per node, shorter distances between nodes, or smaller experimental errors. 

\red{While various quantum repeater protocols address the challenges of long-distance quantum networks, there remains a gap in accurately modeling these protocols for practical implementation. In response, we present a new quantum repeater protocol and analyze it completely by considering all relevant error sources.}

\red{In our novel HEG protocol, qubits in neighboring quantum repeater nodes are entangled with a communication ion in their respective node, which spontaneously emits a time-bin photon whose state depends on the qubit's state. Immediately after emitting the photon, the communication ion becomes disentangled from the rest of the system, leaving the qubit entangled with the emitted photon. The communication ion can therefore be reused with other qubits in the same node to multiplex the HEG in time, increasing the trial rate by up to an order of magnitude. The emitted photons, which are now entangled with the qubits, travel through fibers to a measurement station where a Bell state measurement is performed to herald the entanglement between the qubits. The protocol uses the dipole blockade effect in HEG trials and to perform deterministic entanglement swapping, similar to previous works in both Rydberg atoms \cite{Han2010, Zhao2010} and rare-earth ions \cite{Asadi2018, Debnath2021}. We optimize the protocol performance by considering sessions defined as a fixed number of HEG trials followed by purification and entanglement swapping, similar to Ref. \cite{Munro2010}.} 

The protocol is general and can be used in any quantum system that has a blockade-type interaction, single-photon emitters, and two-qubit gate interactions between the emitters and the qubits. However, in this article we focus on using rare-earth-ion-doped crystals, which are especially attractive to be used as quantum memories or quantum repeaters due to their long coherence times \cite{Zhong2015}, telecom compatibility with erbium ions, potential for integrated photonics, and multimode capacity by leveraging the inhomogeneity in the optical frequencies of the ions \cite{Afzelius2009, Kinos2022a}. For a recent overview of quantum networks using rare-earth ions, see Ref. \cite{Tittel2025}.

We include initialization errors, TQG errors, and qubit measurement errors. Furthermore, instead of assuming a fixed loss due to decoherence as is often done, we consider the exponential decoherence of the qubit memory with time to more accurately model the trade-offs with the other parameters, which is essential when evaluating different implementations of the protocol. We then maximize the achievable secret key rate over a broad range of possible experimental parameters. This allows us to present target values for parameters that must be reached to obtain good secret key rates, thus offering guidance for future quantum technology improvements and innovations.

The remainder of this paper is organized as follows: Section \ref{sec:two_photon_protocol} presents the quantum repeater protocol, where Table \ref{tab:parameters} contains descriptions and symbols of the most important parameters used in this work. Sec. \ref{sec:performance} optimizes the performance of the protocol, and Sec. \ref{sec:experiment} estimates the achievable parameter values for experimental setups. Lastly, we conclude the work in Sec. \ref{sec:conc}.

% ---------------------------------------------------------------
\section{The quantum repeater protocol}\label{sec:two_photon_protocol}
\begin{figure*}
    \centering
    \includegraphics[width=\textwidth]{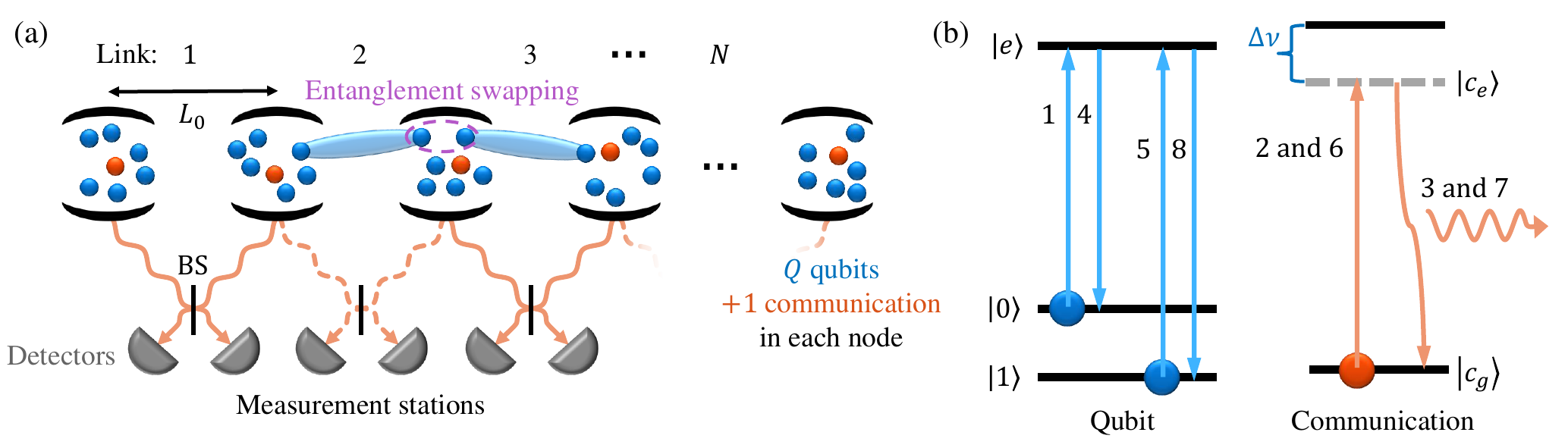}
    \caption{(a) A quantum repeater network with $N$ links, each consisting of two nodes spaced by $L_0$. The links share nodes so that in total $N+1$ nodes exist. Each node has one cavity-coupled communication ion (red dots) and $Q$ qubits (blue dots) that can all block the transition of the communication ion via dipole-dipole interactions. \red{A HEG trial is performed in the following way. (a) A qubit is initialized in a superposition state and the communication ion in its ground state: $\frac{1}{\sqrt{2}}(\ket{0} + \ket{1})\ket{c_g}$. The qubit is then excited on $\ket{0}\rightarrow\ket{e}$ (1), followed by an excitation attempt on the communication ion $\ket{c_g}\rightarrow\ket{c_e}$ (2), which, due to Purcell enhancement, rapidly decays to $\ket{c_g}$ and emits a photon into the cavity (3). Then the qubit is deexcited on $\ket{e}\rightarrow\ket{0}$ (4). After this, the four steps are repeated, except the $\ket{1}\rightarrow\ket{e}$ qubit transition is used (5-8), creating a photon in a later time bin. The dipole blockade prevents the communication ion from being excited if the qubit is excited. Hence, the communication ion is only excited once and emits either an early or a late photon depending on the state of the qubit, i.e., $\frac{1}{\sqrt{2}}(\ket{0}a_L^\dagger + \ket{1}a_E^\dagger)\ket{c_g}$ is obtained, where $a_E^\dagger$ ($a_L^\dagger$) is the creation operator for an early (late) photon. The communication ion is not entangled with the rest of the system and can therefore be reused with a different qubit, i.e., HEG trials can be multiplexed in time as discussed further in the main text. (b) Neighboring nodes simultaneously perform the excitation protocol listed above and transmit the emitted photons via fibers to a measurement station where a beam splitter (BS) and two detectors are used to perform a Bell state measurement. If one early and one late photon is detected the HEG trial is successful and qubit entanglement is heralded into the state described by Eq. (\ref{eq:state_entangled}). In all other cases, the HEG trial fails. When two neighboring links, 2 and 3 in panel (a), each have an entangled pair of qubits, deterministic entanglement swapping (dashed purple ellipse) can be performed on the two qubits in the shared node, thus extending the entanglement.} }
    \label{fig:protocol}
\end{figure*}

\red{A quantum repeater network with $N$ links and $N+1$ nodes is shown in Fig. \ref{fig:protocol}(a). Neighboring nodes are separated by $L_0$, and the end nodes are thus separated by $L_{tot}=NL_0$. Each node consists of a communication ion, used to emit single photons, and $Q$ qubits. The qubits are based on individual rare-earth ions that use nuclear spin states to encode qubit information, while qubit operations are performed via optical transitions. Since all qubits sit within several tens of nanometers they cannot be spatially distinguished. Instead, different qubits are addressed in frequency space, which is possible since each qubit has a unique resonance frequency due to slight variations in the local crystal environment. The transition of the communication ion is Purcell enhanced to increase both the transmission rate and the collection efficiency of emitted photons, which can be achieved using micro-cavities, e.g., photonic crystal cavities \cite{Raha2020, Kindem2020}, open Fabry-P\'{e}rot cavities \cite{Deshmukh2023, Ulanowski2022}, or whispering-gallery-mode cavities \cite{Xia2022}. The qubits must couple strongly to the communication ion via dipole-dipole interactions so that if a qubit is excited the transition of the communication ion is sufficiently frequency-shifted to become off-resonant with any control pulses (dipole-blockaded). For further details regarding how rare-earth qubits work, see Ref. \cite{Kinos2021}.}

\begin{table}
\caption{\label{tab:parameters}Descriptions and symbols of the most important parameters used in this work.}
\begin{ruledtabular}
\begin{tabular}{ll}
\textrm{Description}&\textrm{Symbol}\\
\colrule
Number of links & $N$ \\
Number of qubits per node & $Q$ \\
Link length (distance between neighboring nodes) & $L_0$ \\
Total length (distance between end nodes)  & $L_{tot}$ \\
Fiber attenuation length at telecom wavelengths & $L_{att}$ \\
Fiber-independent efficiency, which includes \\ \quad collection, conversion, and detection efficiencies \\ \quad (not including transmission losses) & $\eta_0$\\
Probability to detect an emitted photon \\ \quad  (including transmission losses) & $\eta$ \\
Difference in phase acquired in the fiber \\ \quad by a late and early photon & $\Delta \phi$ \\
Spin coherence time & $T_2$ \\
Initialization error & $\epsilon_i$ \\
Two-qubit gate error & $\epsilon_{TQG}$ \\
Qubit measurement error & $\epsilon_m$ \\
Round-trip time from node to measurement station & $t_{rt}$ \\
Heralded entanglement generation (HEG) duration & $t_{HEG}$ \\
HEG success probability & $p_{HEG}$ \\
Number of HEG trials per session & $M$ \\
Heralded entanglement purification duration & $t_{pur}$ \\
Rounds of purification at each link  & $P_L$ \\
Rounds of purification between end nodes & $P_E$ \\
Entanglement swapping duration & $t_{swap}$ \\
Session duration & $t_{session}$ \\
EPR generation attempt duration & $t_{EPR}$ \\
EPR generation success probability & $p_{EPR}$ \\
Raw EPR rate & $R$ \\
Secret key rate & $R_{SKR}$ \\
\end{tabular}
\end{ruledtabular}
\end{table}

HEG trials are performed when two neighboring nodes, $A$ and $B$, follow the pulse sequence in Fig. \ref{fig:protocol}(b). As described in the figure caption, for each node this entangles the state of a qubit with the emission of an early or late photon:
\begin{align}\label{eq:state_before_BS}
    \frac{1}{2}(\ket{0}a_L^\dagger + \ket{1}a_E^\dagger)_A \otimes (\ket{0}a_L^\dagger + \ket{1}a_E^\dagger)_B. 
\end{align}
The emitted photons, one from $A$ and one from $B$, are sent through fibers to a beam splitter located at a measurement station, after which the photons are detected using two detectors. Here we assume that the measurement station is in the center between the two nodes, but it can also be integrated into one of the nodes without changing the performance of the protocol. For the HEG to be successful, one early and one late photon must be detected, which heralds the qubits in node $A$ and $B$ into
\begin{align}\label{eq:state_entangled}
    \frac{1}{\sqrt{2}}(\ket{01} \pm e^{i(\Delta\phi_A - \Delta\phi_B)}\ket{10}),
\end{align}
\red{where $+$ ($-$) is used if the photons were detected by the same detector (different detectors)}, and $\Delta\phi_A$ ($\Delta\phi_B$) is the difference in phase acquired in the fiber by a late and an early photon traveling from node $A$ ($B$). The protocol therefore only requires that the optical paths through the individual arms do not change during the time between the early and late photons. This is much more relaxed compared to single-detection HEG protocols, which are sensitive to the optical path difference between the two arms, see, e.g., the discussion in Ref. \cite{Sangouard2011}.

In this work, we assume that the phase differences $\Delta\phi_A$ and $\Delta\phi_B$ are negligible or measurable, the photons from two different nodes are indistinguishable for the single-photon counters used as detectors, and the dark counts of the detectors are sufficiently low, as discussed and motivated in Appendix \ref{app:HEG_assumptions}. Under these assumptions the qubits are prepared into the $\ket{\Psi^+} = \frac{1}{\sqrt{2}}(\ket{01} + \ket{10})$ EPR pair (or $\ket{\Psi^-}$, which is just a local transformation from $\ket{\Psi^+}$). 

The probability that an emitted photon is detected is 
\begin{align}\label{eq:eta_t}
    \eta = \eta_0 e^{-\frac{L_0}{2L_{att}}},
\end{align}
where $L_{att}$ is the attenuation length of the fiber, which for typical fibers at 1.5 $\upmu$m is around 22 km, and $\eta_0$ is the fiber-independent efficiency which contains all other relevant efficiencies: collection efficiency of emitted communication photons, efficiency to convert the communication photons to 1.5 $\upmu$m if needed, and detection efficiency. 

Thus, the success probability for a HEG trial is
\begin{align}\label{eq:p_trial}
    p_{HEG} = \eta^2 / 2,
\end{align}
since we require the detection of two photons and only half of such events register one early and one late photon. 

An important aspect of this protocol is that the communication ion is disentangled from the rest of the system after emitting a photon and can thus be reused together with another qubit to perform a new HEG trial while the photons from previous HEGs are still traveling toward the measurement station for detection. \red{Time multiplexing removes the limitation that HEG trials can only be repeated once every round-trip time, $t_{rt} = L_0/v$ where $v$ is the speed of light in the fiber, i.e., the time until an emitted photon is detected at the measurement station and the heralding information reaches the nodes.} Multiplexing is possible as long as other qubits are available in the nodes since qubits cannot be reused during the round-trip time. However, if a trial fails, the qubits can be reinitialized and reused in future HEG trials once this duration has passed. If qubits are successfully entangled, HEG trials can continue until all available qubits in the nodes have been exhausted. 

\begin{figure}
    \centering
    \includegraphics[width=\columnwidth]{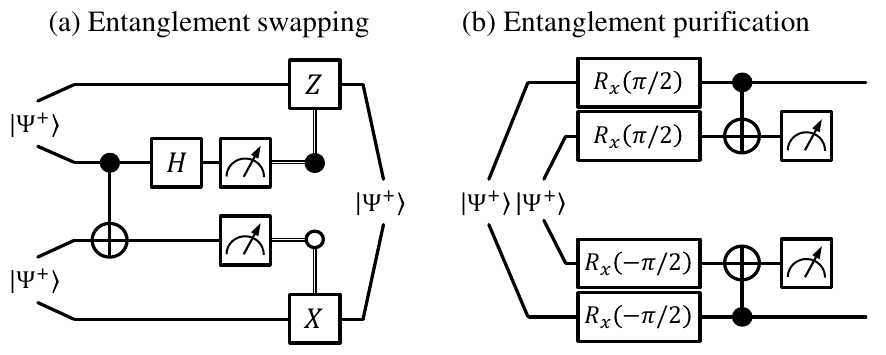}
    \caption{(a) Entanglement swapping protocol \cite{Gottesman1999}: two $\ket{\Psi^+}$ EPR pairs, each having one qubit in the middle node, are deterministically transformed into $\ket{\Psi^+}$ of the two qubits in outer nodes. (b) Heralded entanglement purification protocol \cite{Deutsch1996, Dur1999, Bratzik2013}: two EPR pairs, each having one qubit in the upper node and one qubit in the bottom node, are probabilistically combined into one EPR pair with higher fidelity (outermost lines) if the measurements at the two nodes coincide, otherwise the purification fails and the result is discarded. The success rate is higher if the initial EPR pairs have low errors. The protocol improves the fidelity if there is one bit or one phase error, which are the most likely errors in our systems. Note that other purification protocols might yield better results \cite{Krastanov2019}. For readability, we have separated the single- and two-qubit gates (SQG and TQG) in the two circuits, however, in experiments, the SQGs can be incorporated into the TQGs and measurements similarly as discussed in Sec. IV C of Ref. \cite{Kinos2023}.}
    \label{fig:ES_HEP}
\end{figure}

The network consists of two end nodes and $N-1$ inner nodes, where each inner node is used in two links. Since a communication ion cannot be used simultaneously in two links, HEG trials must alternate between being performed on odd and even links. To achieve this, fiber switches can be used to guide the emitted photons to the appropriate measurement station. These need to be sufficiently fast to switch paths between the emissions of a late photon in one trial and an early photon in the next trial and should ideally have low losses. 

When successful HEG has occurred in two neighboring links, the entanglement distance can be increased by performing deterministic entanglement swapping using the two qubits in the shared node, see Fig. \ref{fig:ES_HEP}(a). When each inner node has performed entanglement swapping, the final EPR pair consists of qubits in the end nodes. 

If the EPR fidelity is ever deemed too low, one can perform heralded entanglement purification and attempt to convert two low-fidelity EPR pairs into one with higher fidelity, see Fig. \ref{fig:ES_HEP}(b). This can be done on EPR pairs connecting arbitrary nodes and not only on EPR pairs between the end nodes. Depending on the connectivity of the qubits within a node, one might need to perform SWAP operations before entanglement swapping or purification. However, this work assumes that the qubits have sufficiently high connectivity such that SWAPs are not necessary, which is reasonable for rare-earth systems \cite{Kinos2022a}. 

% ---------------------------------------------------------------
\section{The performance of the quantum repeater network}\label{sec:performance}
The previous section described the core components of the quantum repeater network, and this section discusses how to best combine these to maximize the rate of high-fidelity EPR pairs between the end nodes. 

One option is to run HEG as often as possible and then perform entanglement swapping and purification according to some rules that maximize the throughput. However, this is difficult to optimize because of the following properties: the success of HEG is random, information only travels at the speed of light, and entangled qubits decohere with time. To show the interplay and trade-offs between performing HEG trials, entanglement swapping, and purification attempts, we now spend some time elaborating on this problem. 

\begin{figure}
    \centering
    \includegraphics[width=\columnwidth]{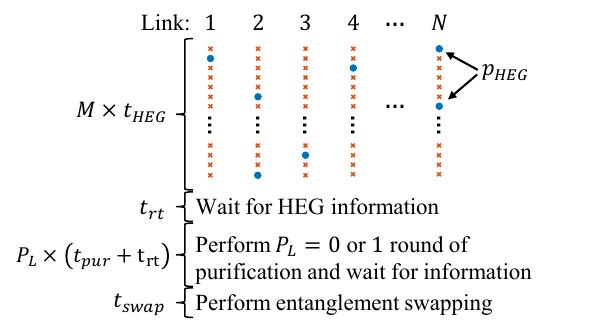}
    \caption{Overview of a session that attempts to generate one EPR pair between the end nodes. It begins with $M$ HEG trials, which have a success probability of $p_{HEG}$ (successes are indicated by blue dots). After the round-trip time between a node and the measurement station, $t_{rt}$, the HEG information is known locally. Following this, either zero or one round of purification is performed on each link with at least three EPR pairs, where the two most recently generated pairs are used in the purification. Then entanglement swapping is performed at each inner node. For successful purification attempts, the purified pair is used in the entanglement swapping, whereas the third most recent pair is used if the purification attempt fails. When no purification is performed in a link the most recently generated pair is used. If any link fails to generate at least one EPR pair, the session fails. Finally, up to two rounds of purification are performed on EPR pairs from sequential sessions to produce one final high-fidelity EPR pair. }
    \label{fig:session}
\end{figure}

To limit errors due to decoherence, entanglement swapping should be performed as soon as possible to reduce the total number of entangled qubits. However, due to the random nature of HEG, one link can have several EPR pairs before an adjacent link obtains its first; so which EPR pair in the former link should be used in the entanglement swapping? Due to decoherence, EPR pairs become worse with time. Thus, using the latest pair results in the highest fidelity. However, if older pairs are never used, the rate of generating useful EPR pairs is reduced. By using purification one can attempt to purify low-fidelity EPR pairs. However, purification decisions are taken at two separate nodes that are not necessarily neighbors, and since information takes time to travel between the nodes it is difficult to construct optimal rules for when to perform purification. Furthermore, purification has a higher success probability if higher fidelity EPR pairs are used, but even a successful purification halves the number of EPR pairs. Thus, there is yet again a trade-off between fidelity and rate. Finally, when entanglement swapping or purification is performed one cannot simultaneously do HEG trials, which again reduces the rate of generated EPR pairs. 

In this work, we optimize the performance of sessions, see Fig. \ref{fig:session}, consisting of a fixed number of $M$ HEG trials, followed by either $P_L = 0$ or $1$ rounds of purification performed on each link, before performing entanglement swapping on all links to create one EPR pair between the end nodes. Note, however, that EPR pairs from sessions can still have too low fidelity and we therefore perform $P_E = 0$, $1$, or $2$ purification steps on EPR pairs from sequential sessions to generate one final high-fidelity EPR pair. The choice of $P_E$ and $P_L$ is taken before running the quantum repeater protocol and is not changed depending on the outcome of a particular session. Thus, we analyze six different protocols consisting of all combinations of $(P_E, P_L)$, and for each protocol we optimize the number of links $N$ and the number of trials per session $M$ to maximize the performance of the network. Analyzing sessions with a fixed duration has, e.g., been done in Ref. \cite{Munro2010}, except they rely on error correction protocols and hence have stricter requirements on low errors from imperfect operations and more qubits per node. 

The probability to generate at least one EPR pair in all $N$ links in a session is given by
\begin{align}\label{eq:p_N_1_M}
    p_{1\rightarrow M}^{(N)} = \left[ 1 - (1 - p_{HEG})^M \right]^N.
\end{align}
Thus, if $M$ is sufficiently large an EPR pair is with high probability generated each session. If desirable, $M$ can be increased even further so that each session generates multiple EPR pairs, and the probabilities for more general cases like this are discussed in Appendix \ref{app:success_rate}, but here we focus on the case where each session tries to generate one EPR pair. 

The duration of a session is 
\begin{align}
    t_{session} = Mt_{HEG} + t_{rt} + P_L(t_{pur}+t_{rt}) + t_{swap},    
\end{align}
where $t_{HEG}$ is the duration of a HEG trial, $t_{rt}$ is the round-trip time from a node to the measurement station where the photons are detected, see Fig. \ref{fig:protocol}(a), $t_{pur}$ and $t_{swap}$ are the durations to perform purification and entanglement swapping, respectively, and we either perform no ($P_L=0$) or one ($P_L=1$) round of purification on each link. 

The Pauli frame of the final EPR pair, i.e., which Bell state it is in, is randomly set by the measurement outcomes of the entanglement swapping steps, and this classical information must travel from each node to the closest end node before purification can be performed on sequential sessions if $P_E \neq 0$. However, we can continue to run sessions while we wait for the entanglement swapping information. Therefore, if all sessions and purification attempts on the end nodes are successful, the time taken to generate one final EPR pair is $t_{EPR} = t_{session}(P_E+1) + t_{pur}P_E$. However, both sessions and purification attempts can fail, so only a fraction $p_{EPR}$ of the attempts succeed, giving a rate to generate EPR pairs of
\begin{align}\label{eq:rate}
    R = \frac{p_{EPR}}{t_{EPR}},
\end{align}
where the calculation of $p_{EPR}$ is given in Appendix \ref{app:success_rate}. For the case where $P_E=0$, this simplifies to 
\begin{align}\label{eq:rate_P_E_0}
    R_{P_E=0} = \frac{p_{1\rightarrow M}^{(N)}}{t_{session}}. 
\end{align}

Another important metric of the quantum repeater network is the error of the final EPR pair. To estimate this we consider the following errors: initialization errors, with probability $\epsilon_{i}$ a qubit is initialized with the wrong phase; TQG errors, modeled as all combinations of two Pauli operator errors, $I$, $X$, $Y$, and $Z$, except $I\otimes I$, so that the total two-qubit error is $\epsilon_{TQG}$; qubit state measurement errors, with probability $\epsilon_m$ a $\ket{0}$ is mistakenly measured as a $\ket{1}$ and vice versa; and decoherence errors, entangled qubits decohere due to the limited spin coherence time, $T_2$. Appendix \ref{app:error_estimation} presents how the errors are calculated analytically. We have also performed numerical simulations to verify the analytical approach, see Appendix \ref{app:numerical_approach}. 

The optimal choice of $M$ and $N$ that maximizes the final EPR rate and fidelity depends on the distance between the end nodes, the error rates, the coherence time, the efficiency $\eta_0$, the time it takes to perform trials, $t_{HEG}$, entanglement swapping, $t_{swap}$, and purification, $t_{pur}$, and the choice of when to perform purification, i.e., the values of $P_L$ and $P_E$. As discussed above, there are several trade-offs between the rate and fidelity, and the optimal $M$ and $N$ depend on how these are weighted against each other. This weighting can depend on the application, e.g., EPR pairs used for the teleportation of gates \cite{Gottesman1999} might have stricter requirements on the successful pair generation rate and the EPR error compared to secret key generation, where one can simply measure the qubits and then wait for the heralded information. 

\begin{figure*}
    \centering
    \includegraphics[width=\textwidth]{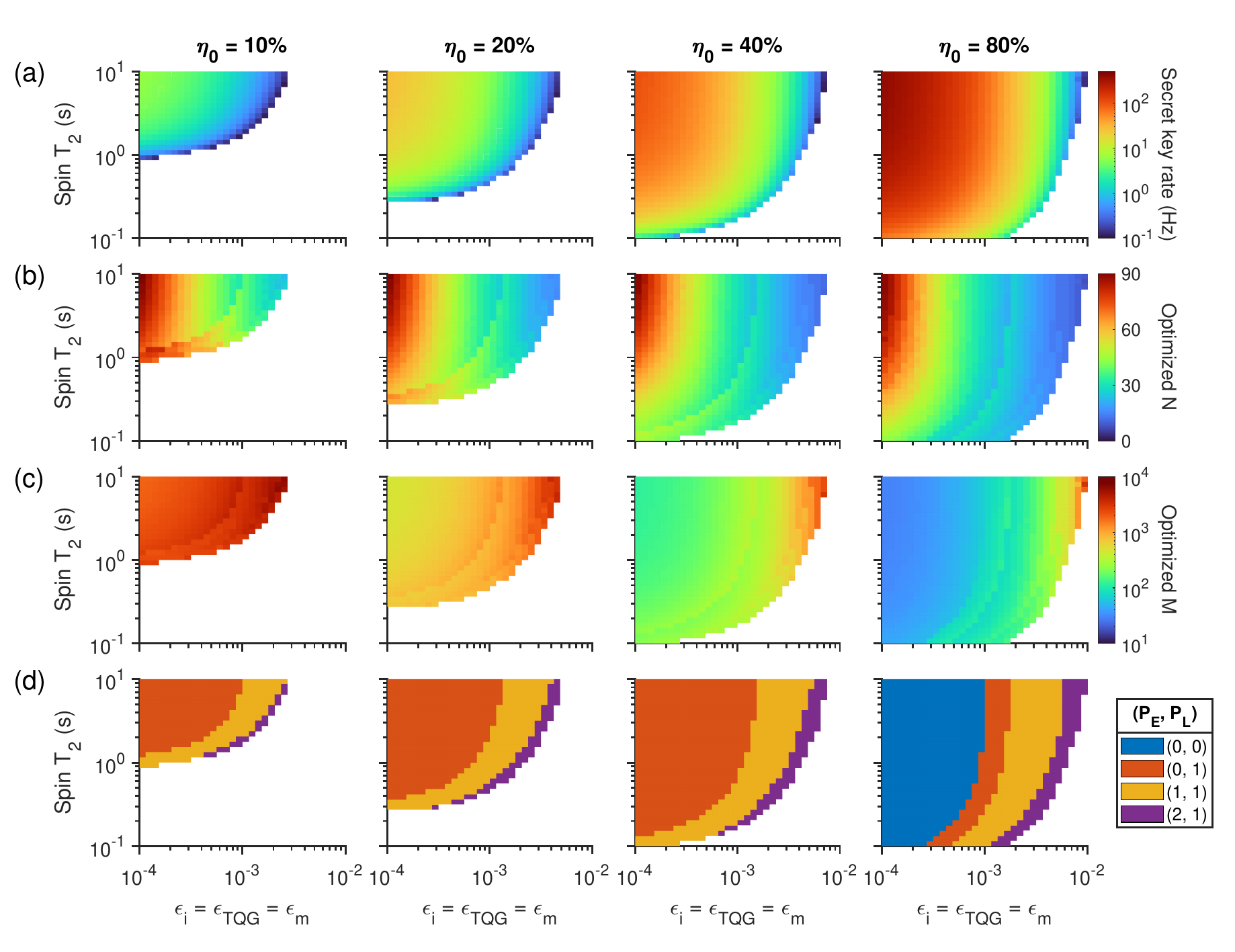}
    \caption{(a) Shows the maximized secret key rate for a quantum repeater network spanning $L_{tot} = 1000$ km, as a function of the error rates, $\epsilon_i = \epsilon_{TQG} = \epsilon_m$ (horizontal axes), and spin coherence time, $T_2$ (vertical axes), for different values of the fiber-independent efficiency $\eta_0$ (columns), which does not include transmission losses through the fiber. White indicates that the rate is lower than $0.1$ Hz. (b) The optimal number of links $N$ ranges from $10$ to $90$, with fewer links being used when the errors are large, when the coherence time is short, or when less purification is used. Hence, the distance between links, $L_0$, varies from around $100$ km down to $11$ km. (c) The optimal number of HEG trials performed on each link in a session, $M$, varies from a few tens to several thousand depending on the parameter values. (d) Shows the optimal purification protocol $(P_E, P_L)$, where $(1, 0)$ and $(2, 0)$ were never the best protocol and are therefore not included. As expected, more purification is needed for high errors, short coherence times, and low efficiencies. $t_{HEG} = 40$ $\upmu$s, $t_{swap} = 210$ $\upmu$s, and $t_{pur} = 220$ $\upmu$s as discussed in Sec. \ref{sec:experiment}. }
    \label{fig:skr}
\end{figure*}

In this work, we consider quantum cryptography between Alice and Bob who are located at the two end nodes, and therefore maximize the secret key rate
\begin{align}\label{eq:rate_SKR}
    R_{SKR} = Rr_{\infty},
\end{align}
where $r_{\infty}$ represents the fraction of secure bits that Alice and Bob can extract from the imperfect EPR pairs, and higher errors give lower $r_{\infty}$. We study the case where Alice and Bob pick measurement basis with biased probability \cite{Lo2005}, i.e., a majority of cases are measured in basis $X$ to extract the key, while a minority of cases are measured in basis $Z$ to estimate the quantum bit error rate, which can be used to determine the degree of error correction and privacy amplification required to reduce the eavesdropper's information about the secret key. We focus on the BB84 quantum key distribution protocol \cite{Bennett2014}, where 
\begin{align}\label{eq:r_inf}
    r_{\infty} = 1 - h(e_Z) - h(e_X),
\end{align}
where $h(p) = -p \log_2(p) - (1-p)\log_2(1-p)$ is the binary entropy and $e_Z$ and $e_X$ are quantum bit errors, for more information see Appendix \ref{app:error_estimation_final} and Ref. \cite{Abruzzo2013}. Note that this provides an upper bound on the secret fraction, which can potentially be reduced by experimental imperfections. 

The maximum secret key rates achieved for a broad range of parameter values are shown in Fig. \ref{fig:skr}(a), and details about the optimization procedure are found in Appendix \ref{app:simulations}. As can be seen, it is possible to achieve secret key rates ranging from a few Hz to several hundred Hz. \red{For each $\eta_0$ there is a limit on how short the coherence time and how high the errors can be before the secret key rate starts decreasing rapidly. This threshold occurs at a rate of approximately $0.1$ Hz. A higher $\eta_0$ makes it easier to achieve a higher rate, thereby allowing for shorter coherence times and larger errors before reaching this threshold.} Decreasing the error is always beneficial, but it is most important to lower the error from $10^{-2}$ toward $10^{-3}$. Similarly, achieving a coherence time of around one second is in most situations enough, and going beyond this has a limited impact except when $\eta_0 \lesssim 10\%$. 

In Figs. \ref{fig:skr}(b)-(c), the optimized values of the number of links $N$ and number of HEG trials per session $M$ can be seen. In general, for a given set of experimental parameters increasing $N$ reduces the distance between nodes, $L_0$, which increases the success probability of HEG trials, see Eqs. (\ref{eq:eta_t}) and (\ref{eq:p_trial}). Thus, $M$ can be reduced while still retaining a large probability to create EPR pairs in all links, see Eq. (\ref{eq:p_N_1_M}). This reduces the session duration, thereby increasing the rate of EPR pair generation and lowering the duration during which qubits decohere. However, increasing $N$ also increases the total number of qubits, which increases the EPR error due to more initialization errors and more qubits that decohere. Furthermore, more entanglement swapping steps are required which increases TQG and measurement errors. 

Thus, if qubit initialization, TQG, and qubit measurement errors are large, $N$ must be low otherwise the EPR errors are too high. However, if these errors are small, then $N$ can be increased to reduce $M$ until either the success probability of HEG trials saturates around $p_{HEG} \rightarrow \eta_0^2/2$ and thus $M$ cannot be reduced any further, or until decoherence errors limit the achievable results due to the addition of more qubits. 

The best purification protocols $(P_E, P_L)$ to use for the different experimental settings are shown in Fig. \ref{fig:skr}(d). In most cases, it is beneficial to use $P_L = 1$, i.e., to perform one round of purification on all links that contain at least three EPR pairs. The reason is that if $M$ is large there is only a small overhead in the session duration to reduce the errors of link EPR pairs. Purification between the end nodes, however, is only beneficial when the final EPR error becomes so large, that despite more than halving the raw EPR rate $R$, the fraction of secure bits $r_\infty$ more than doubles to compensate and increase the secret key rate. 

\begin{figure}
    \centering
    \includegraphics[width=\columnwidth]{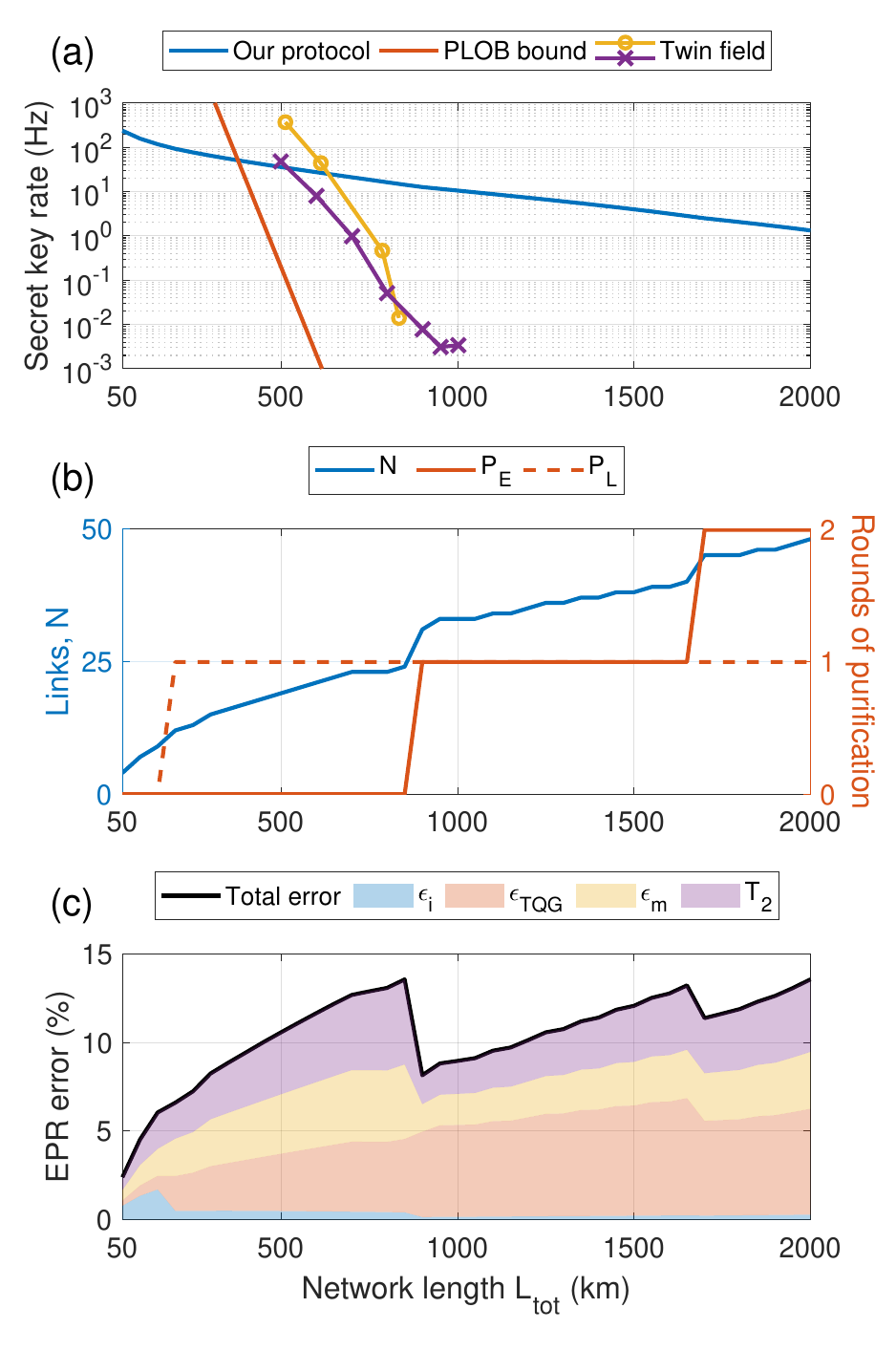}
    \caption{(a) Shows the secret key rate as a function of the network length $L_{tot}$, where $N$ and $M$ have been optimized together with $(P_E, P_L)$ to maximize the rate. For comparison, we show the Pirandola–Laurenza–Ottaviani–Banchi (PLOB) bound \cite{Pirandola2017} which sets the maximum rate that a repeaterless network can achieve when using $L_{att} = 22$ km and a 1 GHz repetition rate. We also show experimental results using the twin field protocol \cite{Lucamarini2018}: from Refs \cite{Wang2022} (yellow circles) and \cite{Liu2023} (purple crosses), both of which used ultra-low-loss optical fibers with $L_{att} \approx 27.5$ km. As can be seen, the PLOB bound and the twin field protocols decrease with a much steeper slope than our protocol. (b) The left axis (blue) shows the optimized number of links $N$ while the right axis (red) shows the purification protocol $(P_E, P_L)$ that gave the maximum rate. (c) The total error of the final EPR pair (black) is shown together with estimations of how initialization errors (blue), TQG errors (red), qubit measurement errors (yellow), and decoherence errors (purple) contribute to the total error. The dips in the EPR error occur because additional purification is used. For all panels, realistic experimental parameters of $\eta_0 = 40\%$, $\epsilon_i = \epsilon_{TQG} = \epsilon_m = 10^{-3}$, and $T_2 = 500$ ms were used together with $t_{HEG} = 40$ $\upmu$s, $t_{swap} = 210$ $\upmu$s, and $t_{pur} = 220$ $\upmu$s, as motivated in Sec. \ref{sec:experiment}. }
    \label{fig:skr_error_N}
\end{figure}

In Fig. \ref{fig:skr_error_N} we analyze the quantum network as a function of its total length, $L_{tot}$ for a set of parameters that we estimate will be achievable experimentally as motivated in Sec. \ref{sec:experiment}. Figure \ref{fig:skr_error_N}(a) compares the secret key rate of our protocol with the maximum achievable rate for repeaterless networks, i.e., the PLOB bound \cite{Pirandola2017}, and experimental results using the twin field protocol \cite{Lucamarini2018}. As can be seen, our protocol is advantageous for longer network distances, whereas for shorter distances the other protocols achieve comparable or even higher rates with much simpler setups. 

\red{There also exists a maximum achievable rate when using repeaters \cite{Pirandola2019}. Since this limit only considers losses in the fiber, it can in principle be used to benchmark the impact of other losses or investigate the relative effectiveness of different protocols. However, the limit states how many bits can be obtained per channel use and must therefore be multiplied by the repetition rate of the system to obtain a secret key rate. Since the repetition rate varies drastically depending on which repeater protocol is used, the maximum achievable secret key rate based on the repeater limit also varies drastically. It is therefore difficult to compare with such a limit. For example, a protocol that is close to the theoretical limit of bits per channel use, might still not be useful if the repetition rate is very low. Furthermore, theoretical protocols often differ in the level of underlying assumptions, encompassing both the extent of optimism regarding achievable experimental parameters and the comprehensiveness of error sources considered within the models. In this work, we therefore focus on the achievable secret key rate with our protocol as shown in Fig. \ref{fig:skr_error_N}(a).}

\red{The optimized number of links and rounds of purification used in our protocol can be seen in Fig. \ref{fig:skr_error_N}(b). If the protocol is implemented, the quantum repeater stations will probably be placed in towns or buildings. Therefore, the lengths of the links will vary, slightly reducing the performance of the protocol. There is also an additional cost of constructing more quantum repeater stations, which we have not considered in our optimization procedure. }

\red{In light of the previous discussions, we believe it would be interesting for a future review article in the field to compare different quantum repeater protocols using similar assumptions and optimism regarding achievable parameters. Other considerations that could also be taken into account are the implementation costs and the real-world layout of repeater stations. }

The secret key rate, Eq. (\ref{eq:rate_SKR}), depends on the error of the final EPR pair through the binary entropy used in Eq. (\ref{eq:r_inf}). Thus, it is important to reduce the error to improve the rate. Figure \ref{fig:skr_error_N}(c) shows estimations of how each error source contributes to the total EPR error. As can be seen, the initialization error is only significant if no purification is used ($P_E = P_L = 0$ for the shortest distances). For longer distances where purification is used, the TQG, measurement, and decoherence errors are of similar magnitudes, and reducing any of them is beneficial for the performance of the protocol. It should be noted that this is not always the case, e.g., if the coherence time increases to $5$ s, its contribution to the EPR error is negligible, and improving it further has no significant impact on the secret key rate. \red{Lastly, increasing $P_E$ significantly reduces the total error, see Fig. \ref{fig:skr_error_N}(b-c), but also reduces the raw EPR rate, $R$. These two effects balance each other so that the secret key rate decreases continuously with the network length as shown in Fig. \ref{fig:skr_error_N}(a).}

\begin{figure}
    \centering
    \includegraphics[width=\columnwidth]{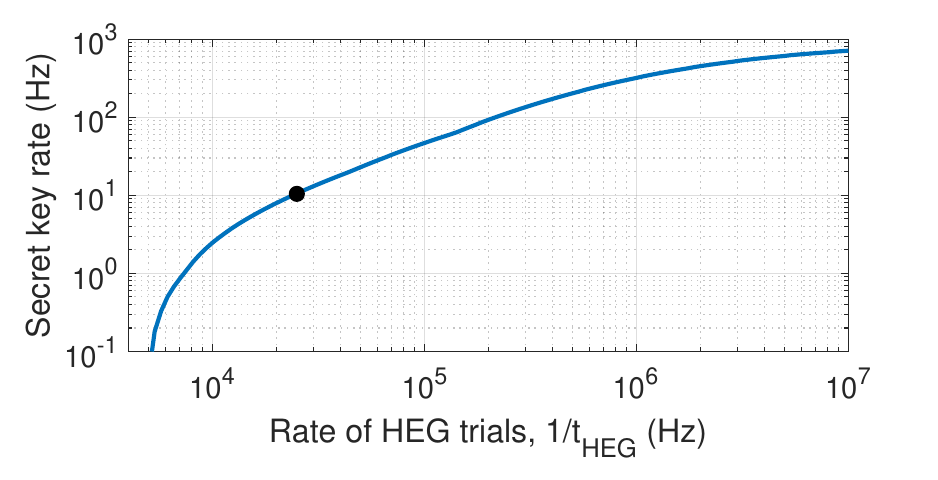}
    \caption{Shows the secret key rate as a function of the HEG trial rate, $1/t_{HEG}$ for a $1000$ km long network. All other parameters are the same as in Fig. \ref{fig:skr_error_N}. The black dot indicates the trial rate of $1/t_{HEG} = \frac{1}{40 \upmu s} = 25$ kHz used in the previous figures.}
    \label{fig:skr_vs_trial}
\end{figure}

So far we have assumed that the rate of HEG trials is $1/t_{HEG} = 25$ kHz, but Fig. \ref{fig:skr_vs_trial} shows how the secret key rate varies as a function of the trial rate. For intermediate trial rates, the secret key rate increases almost linearly. However, for higher trial rates the need to wait for information during the round-trip time, $t_{rt}$, and the time it takes to perform entanglement swapping, $t_{swap} = 210$ $\upmu$s, limits the maximum achievable secret key rate. When the trial rate is low, the time it takes to successfully achieve HEG is comparable to the coherence time and the errors grow rapidly which results in a sharp cutoff in the secret key rate. Thus, for the experimental parameters studied here, it is important to achieve a trial rate larger than around $10$ kHz. \red{Note that it is due to the multiplexing of our protocol that we can assume a HEG trial rate of $25$ kHz. Without multiplexing, the rate is limited by the round-trip time, which alone is $25$ kHz for 8 km long links. Thus, without multiplexing it is difficult to achieve a sufficiently high rate without using many repeater stations.}

\red{Finally, while we in this work focus on optimizing the secret key rate, the protocol’s ability to generate EPR pairs between end nodes makes it valuable for more advanced applications, such as distributed quantum computing.}

% ---------------------------------------------------------------
\section{Estimation of experimental parameter values}\label{sec:experiment}
This section provides estimates for parameter values achievable in an experimental setup that uses rare-earth-ion-doped crystals. We assume that Eu ions doped into a Y$_2$SiO$_5$ crystal at the percent level are used as qubits and Er or Nd ions are used as communication ions, similar to Ref. \cite{Kinos2021}. Single communication ions can, e.g., be detected using a micro-cavity setup, and while combining micro-cavity operation with long coherence times is challenging, it has been achieved using polished-down thin films \cite{Merkel2020}.

Assuming a cavity quality factor of $10^6 \rightarrow 10^7$ and a low mode volume, the reduced lifetime of either Er or Nd could reach around $100$ ns \cite{Kinos2021, Alqedra2023a}. Thus, if single-qubit gate operations, which consist of one excitation and one deexcitation, take in the order of $5$ $\upmu$s \cite{Kinos2021a}, the protocol shown in Fig. \ref{fig:protocol}(b) takes around $10$ $\upmu$s to perform. Since qubits are reused if HEG trials fail, they must repeatedly be reinitialized to $\frac{1}{\sqrt{2}}(\ket{0} + \ket{1})$, which we assume can be done in $10$ $\upmu$s, which might be achievable by a temporary shortening of the qubit lifetime, e.g., by driving from the excited state to another Stark level \cite{Lauritzen2008}. Thus, the time between HEG trials on the same link is $t_{HEG} = 40$ $\upmu$s, since the $20$ $\upmu$s duration must be doubled as we alternate between performing trials on even and odd links. $t_{swap} = 210$ $\upmu$s and $t_{pur} = 220$ $\upmu$s if we assume that a TQG takes $10$ $\upmu$s \cite{Kinos2021a} and a qubit measurement takes $100$ $\upmu$s \cite{Walther2015, Alqedra2023a}. For entanglement swapping the two measurements must be performed sequentially as they are performed on qubits in the same node and therefore use the same communication ion. For purification, both the TQGs and the measurements can be performed in parallel, but each node is involved in two links and therefore needs to perform two purification attempts. 

Each inner node requires the following number of qubits
\begin{align}
    Q = 2(1 + 2P_L + \lceil t_{rt}/t_{HEG} \rceil),
\end{align}
where $1 + 2P_L$ represents how many EPR pairs we need to store to perform purification, and $\lceil t_{rt}/t_{HEG} \rceil$ is how many trials can be performed in the round-trip time, rounded up to make sure that we can always multiplex. Finally, we multiply by $2$ since each inner node is involved in two links. The end nodes only require half this number, plus some additional qubits to store the EPR pairs from previous sessions if $P_E \neq 0$. To achieve the results presented in this work, inner nodes require between $6$ ($L_0 = 11$ km $P_L = 0$) and $32$ ($L_0 = 100$ km $P_L = 1$) qubits, which should be achievable in randomly doped crystals \cite{Kinos2022a}. 

The fiber-independent efficiency $\eta_0$ is harder to estimate since the collection efficiency can range from $10\rightarrow 80\%$ depending on the exact setup and how well the coupling between the cavity and the fiber has been optimized \cite{Hunger2010, Kinos2021}. This value should also include the efficiency of a fast ($\upmu$s time scale) and low-loss fiber switch. Since Er is already at $1.5$ $\upmu$m no wavelength conversion is needed, but for Nd the conversion efficiency could be in the order of $50\%$ \cite{Tanzilli2005, Sangouard2009}. State-of-the-art telecom detectors with low dark counts have an efficiency of $75\rightarrow 98\%$ \cite{Divochiy2008, Krutyanskiy2019, Chang2021}. All combined $\eta_0$ can range from roughly $5 \rightarrow 80\%$. Thus, without going into details about the specific setup used, which is deemed outside the scope of this article, all values of $\eta_0$ shown in Fig. \ref{fig:skr} might be relevant in the future. 

TQG errors of $5\times 10^{-4} \rightarrow 3 \times 10^{-3}$ have been predicted \cite{Kinos2021a}, but depend heavily on the excited state coherence time of the qubits. The initialization error depends on how the reinitialization is implemented, but we assume that it can be done with the same error probability as a TQG. Qubit measurement errors of around $10^{-3}$ seem feasible if buffer ions are used in the measurement process \cite{Walther2015, Alqedra2023a}. 

\red{The spin coherence time is tens of ms \cite{Alexander2007} when running without or with a small magnetic field, around $500$ ms when dynamic decoupling pulses are implemented \cite{Arcangeli2014}, and up to six hours when also operating at the ZEFOZ point, which requires high magnetic fields \cite{Zhong2015}. $500$ ms is in most cases sufficient as shown in Fig. \ref{fig:skr}, and it should therefore be possible to run the quantum repeater protocol without going to the ZEFOZ point. However, the end nodes could benefit from the extended coherence time at the ZEFOZ point, allowing the build-up of a reserve of EPR pairs which can be useful for, e.g., distributed quantum computing.}

All combined, an experimental implementation of the quantum repeater network spanning $1000$ km should be able to reach secret key rates of around $10$ Hz as shown in Fig. \ref{fig:skr_error_N}(a). Note that the rate increases to $\sim 50$ Hz if $\eta_0$ increases to $80 \rightarrow 90\%$ as is often assumed in other works, up to $\sim 90$ Hz if also the coherence time increases to a few seconds, and several hundred Hz if the errors can be made lower. Therefore, future work must perform a more detailed analysis of the exact setup used to obtain better estimates of the achievable rates. 

% ---------------------------------------------------------------
\section{Conclusion}\label{sec:conc}
\red{We have presented a quantum repeater protocol that uses a dedicated communication stage to perform HEG trials at a rate higher than that limited by the travel time between repeater nodes. The HEG trials rely on the detection of two photons, which makes them more robust against phase fluctuations in the arms compared to single-photon protocols \cite{Sangouard2011}. The quantum repeater protocol then uses conventional deterministic entanglement swapping and probabilistic entanglement purification protocols to extend and improve the entanglement. While this work focuses on optimizing secret key rates, the protocol generates EPR pairs and is thus also relevant for more advanced applications such as distributed quantum computing. }

To evaluate the performance of the protocol, we maximize the secret key rate between two end nodes by optimizing the number of links used, how many HEG attempts should be performed before entanglement swapping, and how purification should be implemented. We include initialization errors, TQG errors, qubit measurement errors, and decoherence errors in our analysis. We estimate the rate for a broad range of parameter values, which can help guide future experimental efforts. In summary, it is important to achieve a spin coherence time $\sim 1$ s and to obtain low TQG and measurement errors in the order of $10^{-3}$, while the initialization error is less important if purification is performed. For these parameter values, it is also important that the HEG trial rate is in the order of $10$ kHz. Furthermore, even if the total collection, conversion, and detection efficiency, $\eta_0$ (which does not include the transmission through the fiber), is in the order of $40\%$, a secret key rate of $10$ Hz can be achieved for a network length of $1000$ km, and rates up to several hundred Hz can be achieved if the coherence time, errors, efficiencies, or HEG trial rate improves.

% ---------------------------------------------------------------
\begin{acknowledgments}
We thank Prof. Stefano Pirandola for bringing the theoretical bounds for repeaterless and repeater networks to our attention. This research was supported by Swedish Research Council (no. 2021-03755, and 2019-04949), the Knut and Alice Wallenberg Foundation (KAW 2016.0081), the Wallenberg Center for Quantum Technology (WACQT) funded by The Knut and Alice Wallenberg Foundation (KAW 2021.0009), and the European Metrology Programme for Innovation and Research (20FUN08, NEXTLASERS).
\end{acknowledgments}
% ---------------------------------------------------------------

\appendix
% ---------------------------------------------------------------
\section{Assumptions of the HEG protocol}\label{app:HEG_assumptions}
This section discusses and motivates three assumptions needed for HEG trials to produce pure $\ket{\Psi^+}$ states: that $\Delta\phi_A$ and $\Delta\phi_B$ in Eq. (\ref{eq:state_entangled}) are either zero or can be measured; that photons emitted from different nodes are indistinguishable for the single-photon counters used as detectors; and that the detector dark count rates are negligible. 

$\Delta\phi_A$ is the phase difference acquired by a late and an early photon traveling from node $A$ (equivalent for $\Delta\phi_B$). Using the estimated time scales presented in Sec. \ref{sec:experiment}, it takes $10$ $\upmu$s to perform the protocol shown in Fig. \ref{fig:protocol}(b) and the time between a late and an early photon is in the order of $\Delta t = 5$ $\upmu$s. Since only the difference in phase between the two photons is interesting, slow variations in phase, i.e., phase noise at frequencies lower than around $1/(5$ $\upmu$s$) = 200$ kHz, do not contribute. If the communication ions have a Purcell-enhanced lifetime of $100$ ns, the spontaneously emitted photons have a bandwidth in the order of $10$ MHz. For the results presented in Fig. \ref{fig:skr} the distance from a node to the measurement station, $L_0 / 2$, lies somewhere in the range of $5.5 \rightarrow 50$ km, and at these lengths we assume that the integrated fiber phase noise between roughly $200$ kHz and $10$ MHz is sufficiently low to be negligible. Furthermore, even if for some reason $\Delta\phi_A$ would be too large, it is sufficient to measure and know this phase difference to run the HEG protocol, i.e., the protocol does not require phase-stabilized arms. 

For the photons from nodes $A$ and $B$ to be considered indistinguishable, they must arrive at the beam splitter simultaneously and with the same frequency. Here we assume that the protocol timings and distances to the measurement station can be synchronized so that the photons arrive simultaneously. For the photons to be indistinguishable in frequency, the transition frequencies of communication ions in neighboring nodes have to be well within the frequency bandwidth of the emitted photons, i.e., less than $\sim 10$ MHz. This can potentially be achieved by applying electric fields to shift the transitions of communication ions \cite{Huang2023}, or in the frequency conversion process \cite{Ates2012}. However, since single-photon counters, which are insensitive toward photon phases, are used as detectors, the scheme is robust against known phase and frequency variations of the photons \cite{Debnath2021}. Thus we only require the photons to be indistinguishable with respect to their timing. 

Note that even if $\Delta\phi_A \neq 0$ or the photons are not completely indistinguishable, the state achieved after performing HEG would be a mixed state of $\ket{\Psi^+}$ and $\ket{\Psi^-}$, which is the same type of error as the initialization error, $\epsilon_i$, and can thus be modeled in a similar way. Furthermore, as shown in Fig. \ref{fig:skr_error_N}(c), such errors do not significantly impact the final results as long as some purification is performed. 

Detector dark counts can result in HEG producing the desired detections at the detectors even if the atomic state is wrong. For example, for $\ket{11}$, which only produces two early photons, HEG can still be marked successful if one early photon is detected and one dark count occurs when a late photon should have arrived, i.e., in the detection window of a few lifetimes, $\sim 100$ ns, of the communication ions. Such events become significant when the probability of this occurring is similar to the probability that an emitted photon is detected, $\eta$. We analyze $\eta_0 \geq 10\%$ and from the results of Fig. \ref{fig:skr}, $L_0 \leq 100$ km, which gives $\eta \geq 0.01$. Therefore, the dark count rate should be a small fraction $\epsilon$ of $\sim \eta / 100$ ns $\approx 100$ kHz for the dark count error to be in the order of $\epsilon$. Since high-efficiency superconducting single-photon counters can achieve dark count rates in the order of $1$ Hz, this error is assumed to be negligible. 

% ---------------------------------------------------------------
\section{The success probability of generating EPR pairs}\label{app:success_rate}
The probability that HEG in one link succeeds $k$ times is
\begin{align}\label{eq:p_1_k}
    p_k^{(1)} = \binom{M}{k} p_{HEG}^k (1-p_{HEG})^{M-k},
\end{align}
where $p_{HEG}$ given by Eq. (\ref{eq:p_trial}) is the success probability of a HEG trial. Furthermore, the probability that one link succeeds at least $k$ times is given by
\begin{align}\label{eq:p_1_k_to_M}
    p_{k\rightarrow M}^{(1)} = 1 - \sum_{j=0}^{k-1} p_j^{(1)}.
\end{align}
The probability for $N$ links to succeed at least $k$ times is
\begin{align}\label{eq:p_N_k_to_M}
    p_{k\rightarrow M}^{(N)} = \left( p_{k\rightarrow M}^{(1)} \right)^N,
\end{align}
which is the important probability when designing a session that should generate $k$ EPR pairs. For analysis purposes, it is also interesting to calculate the probability that $N$ links generate at least $k$ EPR pairs, but at least one link does not generate $k+1$ pairs;
\begin{align}\label{eq:p_N_k}
    p_{k}^{(N)} = p_{k\rightarrow M}^{(N)} - p_{k+1\rightarrow M}^{(N)}. 
\end{align}

When $P_E$ rounds of purification are performed on sequential sessions, one attempt to produce one EPR pair takes $t_{EPR} = t_{session}(P_E+1) + t_{pur}P_E$. However, only a fraction $p_{EPR}$ of the attempts succeed, which we estimate by calculating the probability that an attempt succeeds multiplied by how many sessions are used in this case, divided by $p_{sum}$ which is the expectation value of how many sessions are used before we succeed: 
\begin{align}
    p_{EPR} &= \frac{(P_E+1)p^{P_E+1}\prod_{h=1}^{P_E}p_{P_E}^{(h)}}{p_{sum}}, \nonumber\\
    p_{sum} &= (P_E+1)p^{P_E+1}\prod_{h=1}^{P_E}p_{P_E}^{(h)} \nonumber\\
    &+ \sum_{s=1}^{P_E+1} s p^{s-1}(1-p) \prod_{h=1}^{s-2}p_{P_E}^{(h)} \nonumber\\
    &+ \sum_{h=1}^{P_E} (h+1)p^{h+1} (1 - p_{P_E}^{(h)}) \prod_{h'=1}^{h-1}p_{P_E}^{(h')}, 
\end{align}
where we use $p = p_{1\rightarrow M}^{(N)}$ to simplify the expression, and $p_{P_E}^{(h)}$ is the probability that purification number $h$ between the end nodes is successful, which depend on the errors of the two states that are combined and is described in Appendix \ref{app:error_estimation}. 

% ---------------------------------------------------------------
\section{EPR pair error estimation}\label{app:error_estimation}
Here we study the error of the final EPR pair between the end nodes, which includes initialization errors, TQG errors, qubit measurement errors, and decoherence errors. \red{The error is characterized in the Bell state basis, where the system's state is expressed as a combination of the four Bell states}
\red{\begin{align}\label{eq:Bell_states}
    \ket{\Phi^+} &= \frac{\ket{00} + \ket{11}}{\sqrt{2}}, \nonumber\\
    \ket{\Psi^-} &= \frac{\ket{01} - \ket{10}}{\sqrt{2}}, \nonumber\\
    \ket{\Psi^+} &= \frac{\ket{01} + \ket{10}}{\sqrt{2}}, \nonumber\\
    \ket{\Phi^-} &= \frac{\ket{00} - \ket{11}}{\sqrt{2}}. 
\end{align}}
\red{The coefficients $A$, $B$, $C$, and $D$ represent the probabilities of the system being in each respective Bell state, with their sum constrained to $1$. By analyzing these probabilities, the nature and extent of the error can be quantified.} Refs. \cite{Renner2005, Kraus2005} have proved that any arbitrary two-qubit state is just a local twirling operation away from the diagonal Bell states, which is why we can analyze the errors in this basis without compromising the security of the protocol \cite{Abruzzo2013}.

We start by analyzing the Bell state coefficients of an EPR pair from a single session, either without purification ($P_L = 0$) or with one round of purification ($P_L = 1$), then we see how this is modified when purification is used between EPR pairs in the end nodes ($P_E \neq 0$). Here we present our theoretical estimations of the error, but we have also performed numerical studies to validate that our theoretical approach is correct, which are explained in Appendix \ref{app:numerical_approach}.

% ---------------------------------------------------------------
\subsection{EPR pair from a single session without purification}\label{app:session_no_HEP}
The ideal state after HEG has $C = 1$, but with a probability $\epsilon_{i}$ one of the two qubits in the entangled pair has a phase error, i.e., if one error occurs $C$ changes to $B$, but if zero or two errors occur we still get $C$. With $N$ links the state becomes
\begin{align}\label{eq:err_init}
    A_{i} &= 0, \nonumber \\
    B_{i} &= \sum_{n=1}^{N} \epsilon_{i}^{2n-1} (1 - \epsilon_{i})^{2N-2n+1}\binom{2N}{2n-1}, \nonumber \\
    C_{i} &= \sum_{n=0}^{N} \epsilon_{i}^{2n} (1 - \epsilon_{i})^{2N-2n}\binom{2N}{2n}, \nonumber \\
    D_{i} &= 0.
\end{align}

When purification is not used, the links are directly combined using $N$-1 entanglement swapping steps, which can introduce TQG and measurement errors. A TQG error changes one Bell state into any of the other three states with probability $\epsilon_{TQG}/3$. This can be estimated using
\begin{align}\label{eq:err_TQG}
    X_{TQG} &= X_{i} + (X_{i} - \frac{1}{4}) \nonumber \\
    &\times \sum_{n=1}^{N-1} \left(-\frac{4\epsilon_{TQG}}{3}\right)^n \binom{N-1}{n}, 
\end{align}
where $X$ is any of the four Bell state coefficients. 

The entanglement swapping step results in two bits of classical information that determine the Pauli frame of the output state, which can either be used directly to turn the output into $\ket{\Psi^+}$ as shown in Fig. \ref{fig:ES_HEP}(a), or transmitted to the end nodes and combined with the information from all $N$-1 entanglement swapping steps to determine the Pauli frame of the final EPR pair. However, an error in the measurement provides the wrong classical information, resulting in either a bit flip or phase flip if one measurement error occurs, or both a bit and phase flip if two measurement errors occur. With $N$ links and a measurement error of $\epsilon_m$ this can be estimated using
\begin{align}\label{eq:err_readout}
    X_{m} &= X_{TQG} + \frac{X_{TQG}+Y_{TQG}-1/2}{2} \nonumber\\
    &\times \sum_{n=1}^{2(N-1)} \left(-2\epsilon_m\right)^n \binom{2(N-1)}{n} \nonumber\\
    &+ \frac{X_{TQG}-Y_{TQG}}{2} \sum_{n=1}^{N-1} \left(-2\epsilon_m\right)^n \binom{N-1}{n}, 
\end{align}
where $Y$ is the Bell state coefficient that differs in both a bit and phase flip compared to $X$, i.e., for $X = A/B/C/D$ then $Y = B/A/D/C$. 

The entangled qubits also decohere due to the limited spin coherence time, $T_2$, and that error is estimated as 
\begin{align}\label{eq:err_T2}
    X_{session} &= \frac{X_{m} + Z_{m}}{2} + \frac{X_{m} - Z_{m}}{2} e^{-t_{q, tot}/T_2},
\end{align}
where $Z$ is the Bell state coefficient that differs in a phase flip compared to $X$, i.e., for $X = A/B/C/D$ then $Z = D/C/B/A$, and $t_{wait}$ is the total cumulative time the qubits have been entangled. When purification is not used and all links generate at least one EPR pair, then from Fig. \ref{fig:session} we can see that $t_{wait} = 2(mt_{HEG} + N(t_{rt}+t_{swap}))$, where we multiply by $2$ since each link consists of two qubits, and $m$ is the cumulative number of trials all links failed after their last successful HEG, which can be any number between $0$, if all links succeeded on their last trial, to $N(M-1)$, if all links only succeeded on their first trial. The probability to get $m$ when using $N$ links and $M$ trials per session is
\begin{align}\label{eq:p_m_link_tot}
    p(m) &= \frac{p_{HEG}^N}{p_{1\rightarrow M}^{(N)}}(1-p_{HEG})^{m} c^{(N)}(m),
\end{align}
where $c^{(N)}(m)$ is the number of possible ways $m$ can be obtained, which is calculated using the recursive formula
\begin{align}
    c^{(1)}(m) &= 1, \nonumber \\
    c^{(N)}(m) &= \sum_{k=\textrm{max}[m-(N-1)(M-1), 0]}^{\textrm{min}[M-1, m]} c^{(N-1)}(m - k).
\end{align}

% ---------------------------------------------------------------
\subsection{EPR pair from a single session with one round of purification}\label{app:session_HEP}
Purification requires at least two EPR pairs in a link to be performed. However, if only two pairs exist and purification fails, the entire session fails. Thus, to not affect the session success rate, purification is only attempted if at least three EPR pairs are generated in a link. In principle, we have three options on how to perform purification: using the \nth{1} and \nth{2} to last successful trials in purification and using the \nth{3} successful trial as a reserve in case the purification fails, using the \nth{1} and \nth{3} with the \nth{2} in reserve, or using the \nth{2} and \nth{3} with the \nth{1} in reserve. In this work, we only consider the first option. 

If a link does not have enough EPR pairs, purification cannot be performed and the last EPR pair generated is used when performing entanglement swapping. The probability that purification can be performed is given by $p_{pur} = p_{3\rightarrow M}^{(1)}/p_{1\rightarrow M}^{(1)}$, where we normalize with the probability that at least one EPR pair is generated in the link since we are only interested in the error of successful sessions. 

To estimate the final EPR pair error we must estimate the Bell state coefficients for a single link before combining several links using entanglement swapping. For each link, we have three outcomes: purification cannot be performed, purification can be performed but failed, and purification can be performed and succeeded. 

If we cannot perform purification the Bell state coefficients of a single link are estimated using Eq. (\ref{eq:err_init}) with $N=1$ and then adding decoherence as in Eq. (\ref{eq:err_T2}), but using $X_i$ and $Z_i$, and $t_{wait}$ is replaced with 
\begin{align}
    t_{no \ pur} = 2(m t_{HEG} + 2t_{rt} + t_{pur} + t_{swap}),
\end{align}
where $m = 0 \rightarrow M-1$ is the number of trials that occurred after the last successful trial, given the knowledge that the link did not generate at least the three EPR pairs needed to perform purification. The probability to obtain $m$ is
\begin{align}
    p(m) &= p_{HEG} (1-p_{HEG})^{M-1} \nonumber\\
    &+ p_{HEG}^2 (1-p_{HEG})^{M-2} (M-m-1), \nonumber \\
    p(m) &= \frac{p(m)}{\sum_{m=0}^{M-1} p(m)},
\end{align}
where we normalize the probability to sum to $1$. When performing the optimization of the quantum repeater protocols, we only use the expectation value of the duration $\langle t_{no \ pur} \rangle$, which we have verified gives a good agreement compared to using the full distribution. 

If purification was attempted but failed, the third last successful trial is used. The same estimation as above can be used with $t_{res} = t_{no \ pur}$ except $m = 2 \rightarrow M-1$ is now the number of trials that occurred after the third successful trial, and the probability to obtain $m$ is now given by
\begin{align}
    p(m) = \frac{p_{HEG}^3 (1-p_{HEG})^{m-2}}{p_{3\rightarrow M}^{(1)}} \binom{m}{2}.
\end{align}
As above we only use $\langle t_{res}\rangle$ in the optimization. 

For the cases where purification succeeds, we start by estimating the error of two EPR pairs before purification is performed using the same steps as above but with the duration
\begin{align}
    t_{pur} = 2(m_i t_{HEG} + t_{rt}), 
\end{align}
where $m_i$ is either the number of trials after the \nth{1} or the \nth{2} last successful trial, which have a joint probability distribution given by
\begin{align}
    p(m_1, m_2) = \frac{p_{HEG}^2 (1-p_{HEG})^{m_2-1}}{p_{3\rightarrow M}^{(1)}} \nonumber\\
    \times \left(1 - (1-p_{HEG})^{M-m_2-1}\right),
\end{align}
where $m_2 = 1\rightarrow M-2$ and $m_1 = 0 \rightarrow m_2-1$. In optimization, we use the expectation value $\langle t_{pur}\rangle$ for both pairs. 

Purification is then performed according to Fig. \ref{fig:ES_HEP}(b) and Ref. \cite{Deutsch1996}, and therefore includes TQG errors, which we model by first following Eq. (\ref{eq:err_TQG}) with $N=1$ and where $X_i$ is replaced with the coefficients obtained after following the steps for the two EPR pairs described above. The resulting $X_{1}$ and $X_{2}$ are then combined using 
\begin{align}\label{eq:HEP_suc}
    p_s &= (A_1+B_1) (A_2+B_2) + (C_1+D_1) (C_2+D_2), \nonumber \\
    A_s &= \frac{1}{p_s} \Bigl[A_1 A_2 + B_1 B_2 + \frac{\epsilon_{TQG}}{3} \nonumber \\
    &\times \bigl(-(A_1+B_1) (A_2+B_2) + (C_1+D_1) (C_2+D_2)\bigr)\Bigr],  \nonumber \\
    B_s &= \frac{1}{p_s} \Bigl[C_1 D_2 + D_1 C_2 + \frac{\epsilon_{TQG}}{3}  \nonumber \\
    &\times \bigl(+(A_1+B_1) (A_2+B_2) - (C_1+D_1) (C_2+D_2)\bigr)\Bigr], \nonumber \\
    C_s &= \frac{1}{p_s} \Bigl[C_1 C_2 + D_1 D_2 + \frac{\epsilon_{TQG}}{3}  \nonumber \\
    &\times \bigl(+(A_1+B_1) (A_2+B_2) - (C_1+D_1) (C_2+D_2)\bigr)\Bigr], \nonumber \\
    D_s &= \frac{1}{p_s} \Bigl[A_1 B_2 + B_1 A_2 + \frac{\epsilon_{TQG}}{3}  \nonumber \\
    &\times \bigl(-(A_1+B_1) (A_2+B_2) + (C_1+D_1) (C_2+D_2)\bigr)\Bigr], 
\end{align}
where $p_s$ is the probability that the purification is successful. 

To add measurement errors, we also need to calculate the Bell state coefficients obtained if purification fails:
\begin{align}\label{eq:HEP_fail}
    p_f &= (A_1+B_1) (C_2+D_2) + (C_1+D_1) (A_2+B_2), \nonumber \\
    A_f &= \frac{1}{p_f} \Bigl[A_1 C_2 + B_1 D_2 + \frac{\epsilon_{TQG}}{3} \nonumber \\
    &\times \bigl(-(A_1+B_1) (C_2+D_2) + (C_1+D_1) (A_2+B_2)\bigr)\Bigr],  \nonumber \\
    B_f &= \frac{1}{p_f} \Bigl[D_1 A_2 + C_1 B_2 + \frac{\epsilon_{TQG}}{3}  \nonumber \\
    &\times \bigl(+(A_1+B_1) (C_2+D_2) - (C_1+D_1) (A_2+B_2)\bigr)\Bigr],  \nonumber \\
    C_f &= \frac{1}{p_f} \Bigl[C_1 A_2 + D_1 B_2 + \frac{\epsilon_{TQG}}{3}  \nonumber \\
    &\times \bigl(+(A_1+B_1) (C_2+D_2) - (C_1+D_1) (A_2+B_2)\bigr)\Bigr],  \nonumber \\
    D_f &= \frac{1}{p_f} \Bigl[A_1 D_2 + B_1 C_2 + \frac{\epsilon_{TQG}}{3}  \nonumber \\
    &\times \bigl(-(A_1+B_1) (C_2+D_2) + (C_1+D_1) (A_2+B_2)\bigr)\Bigr],  \nonumber \\
\end{align}
where $p_f$ is the probability that the purification fails. Thus, the state directly after performing purification is
\begin{align}\label{eq:HEP_m}
    p_{m} &= \bigl((1-\epsilon_m)^2 + \epsilon_m^2\bigr)p_s + 2(1-\epsilon_m)\epsilon_m p_f\nonumber \\
    A_{m} &= \frac{1}{p_{m}} \Bigl[\bigl((1-\epsilon_m)^2 + \epsilon_m^2\bigr)p_s A_s \nonumber \\
    &+ 2(1-\epsilon_m)\epsilon_m p_f A_f \Bigr],  \nonumber \\
    B_{m} &= \frac{1}{p_{m}} \Bigl[\bigl((1-\epsilon_m)^2 + \epsilon_m^2\bigr)p_s B_s \nonumber \\
    &+ 2(1-\epsilon_m)\epsilon_m p_f B_f \Bigr],  \nonumber \\
    C_{m} &= \frac{1}{p_{m}} \Bigl[\bigl((1-\epsilon_m)^2 + \epsilon_m^2\bigr)p_s C_s \nonumber \\
    &+ 2(1-\epsilon_m)\epsilon_m p_f C_f \Bigr],  \nonumber \\
    D_{m} &= \frac{1}{p_{m}} \Bigl[\bigl((1-\epsilon_m)^2 + \epsilon_m^2\bigr)p_s D_s \nonumber \\
    &+ 2(1-\epsilon_m)\epsilon_m p_f D_f \Bigr],  \nonumber \\
\end{align}
where $p_{m}$ is the probability that we measure that the purification is successful. 

After the purification, the state decoheres according to Eq. (\ref{eq:err_T2}) with the duration $2(t_{pur}+t_{rt}+t_{swap})$. 

To summarize, with probability $1 - p_{pur}$ purification is never attempted and the last successful trial is used, with probability $p_{pur} (1-p_m)$ purification is attempted but failed and the reserved EPR pair is used instead, and with probability $p_{pur} p_m$ purification is attempted and succeeded and the Bell state coefficients for a single link are given by Eq. (\ref{eq:HEP_m}).

Now, $N$ links are combined using $N$-1 entanglement swapping steps that combine the Bell state coefficients of the left, $X_L$, and right, $X_R$, links using the following formulas in an iterative process
\begin{align}
    A_{swap} &= A_L C_R + C_L A_R + B_L D_R + D_L B_R,  \nonumber \\
    B_{swap} &= A_L D_R + D_L A_R + B_L C_R + C_L B_R,  \nonumber \\
    C_{swap} &= A_L A_R + B_L B_R + C_L C_R + D_L D_R,  \nonumber \\
    D_{swap} &= A_L B_R + B_L A_R + C_L D_R + D_L C_R.
\end{align}
Since each link consists of three different sets of Bell state coefficients (no purification, purification succeeded, and purification failed and reserve is used), the number of sets grows exponentially with $N$ when combining several links using entanglement swapping. When optimizing the repeater protocols, we keep the number of sets below $100$ by using a k-means algorithm where the distances are weighted with the probability of obtaining a set. 

Lastly, we add TQG and measurement errors from the entanglement swapping using Eqs. (\ref{eq:err_TQG}) and (\ref{eq:err_readout}) which gives us $X_{session}$ when running the protocol with $P_L = 1$.

% ---------------------------------------------------------------
\subsection{EPR pair between end nodes}\label{app:error_estimation_final}
If no purification is used at the end nodes ($P_E = 0$), then the average fidelity of the EPR pair is given by $\langle C\rangle$ and the quantum bit errors $e_X$ and $e_Z$ used in Eq. (\ref{eq:r_inf}) can be calculated using \cite{Abruzzo2013} 
\begin{align} \label{eq:quantum_bit_error}
    e_X = \langle B\rangle + \langle D\rangle, \nonumber \\
    e_Z = \langle A\rangle + \langle D\rangle,
\end{align}
since $C$ is our desired state. 

When $P_E \neq 0$, we need to combine $P_E + 1$ sequential sessions into one final EPR pair using purification, using $X_{session}$ from either Sec. \ref{app:session_no_HEP} when $P_L = 0$ or Sec. \ref{app:session_HEP} when $P_L = 1$. 

For the first round of purification, decoherence errors are added according to Eq. (\ref{eq:err_T2}) with the duration $t_{session} + L_{tot}/(2v)$ for the first EPR pair and $L_{tot}/(2v)$ for the second, since we need to wait for the HEG information to reach the closest end node to know the Pauli frame of the EPR pairs before purification can be performed. Then we add TQG errors according to Eq. (\ref{eq:err_TQG}) to each pair before using Eqs. (\ref{eq:HEP_suc})-(\ref{eq:HEP_m}) to perform the purification. Finally, we let the state decohere for a time $t_{pur}$. Since we discard the results if purification cannot be performed due to session failures or if purification fails, we are now only interested in the results when purification succeeds, but $p_{m}$ in Eq. (\ref{eq:HEP_m}) is still used to calculate $p_{P_E}^{(h)}$ in Appendix \ref{app:success_rate}. 

If $P_E > 1$, the results of the previous purification are combined with a new session using the same steps as above, except the previous results only decohere during an additional time of $t_{session}$ instead of $t_{session} + L_{tot}/(2v)$. The information regarding the success of the first purification arrives at the end nodes after $L_{tot}/v \approx 5$ ms for a $1000$ km long network, and in the cases where $P_E > 1$, $t_{session} > 5$ ms so this information has time to arrive before the second round of purification is attempted. 

Finally, the quantum bit errors can be estimated using Eq. (\ref{eq:quantum_bit_error}). 

% ---------------------------------------------------------------
\section{Optimizing the secret key rate}\label{app:simulations}
For each combination of the purification protocols $P_L = 0\rightarrow 1$ and $P_E = 0\rightarrow 2$, we optimize $N$ and $M$ to achieve the highest secret key rate given in Eq. (\ref{eq:rate_SKR}). This is done using MATLAB's patternsearch function, which we run with the following options: caches are used, GPSPositiveBasis2N is the poll method, scale mesh is off, accelerate mesh is false, and TolMesh is 0.9. The raw EPR rate, $R$, and the EPR quantum bit errors, $e_X$ and $e_Z$, are calculated as described in Appendices \ref{app:success_rate} and \ref{app:error_estimation}. 

% ---------------------------------------------------------------
\section{Numerical validation of the theoretical calculations} \label{app:numerical_approach}
\red{To validate the theoretical calculations, we also perform numerical simulations to estimate the rate and average fidelity of the final EPR pairs. We validate the theory for a wide range of parameters and settings, and for each we simulate 1000 sessions to gather enough statistics to see an agreement with the theory. }

\red{Each session is simulated by randomizing which of the $M$ HEG trials succeed for each of the $N$ links. If all HEG trials fail for at least one link, the session fails. Otherwise, we define a 4x4 density matrix for each link that describes the state of the two qubits in the generated EPR pair. These start in a pure Bell state, but then initialization, TQG, measurement, and decoherence errors are added whenever HEP or entanglement swapping is performed, or time passes. Errors are added based on the same principles as outlined in Appendix \ref{app:error_estimation}. However, since we now simulate which HEG trials are successful, the exact duration that each EPR pair has decohered is calculated and used when adding decoherence errors. }

\red{If $P_L \neq 0$, then HEP is attempted on all links containing at least three EPR pairs. The outcome is randomized, and if successful, the two most recently generated EPR pairs are combined into one EPR pair with higher fidelity. If HEP fails, the third most recently generated EPR pair is used instead, which has worse fidelity due to increased decoherence. }

\red{After this, entanglement swapping combines the density matrices for all links, two at a time. This generates a final density matrix for the EPR pair of that session. If $P_E = 0$, this is used to calculate the error of the final EPR pair. If $P_E \neq 0$, more sessions are simulated, and if sufficiently many are successful in a row, HEP is attempted. If unsuccessful, no final EPR pair is generated, thus reducing the rate. If successful, the density matrices are combined into one with higher fidelity, which is used to estimate the final EPR error. }

\red{The rate of generating final EPR pairs is determined by checking how many pairs were successfully created during the time it takes to perform 1000 sessions. In conclusion, the numerical simulations agreed with the theory within the statistical uncertainty for all cases that were tested, indicating that the theory is valid.}

% ---------------------------------------------------------------
\bibliography{Ref_lib}% Produces the bibliography via BibTeX.

\end{document}